\documentclass[sigconf]{acmart}
\AtBeginDocument{%
  \providecommand\BibTeX{{%
    \normalfont B\kern-0.5em{\scshape i\kern-0.25em b}\kern-0.8em\TeX}}}

\copyrightyear{2023} 
\acmYear{2023} 
\setcopyright{acmlicensed}\acmConference[CHI '23]{Proceedings of the 2023 CHI Conference on Human Factors in Computing Systems}{April 23--28, 2023}{Hamburg, Germany}
\acmBooktitle{Proceedings of the 2023 CHI Conference on Human Factors in Computing Systems (CHI '23), April 23--28, 2023, Hamburg, Germany}
\acmPrice{15.00}
\acmDOI{10.1145/3544548.3581512}
\acmISBN{978-1-4503-9421-5/23/04}

\usepackage{xcolor}
\usepackage{enumitem}
\usepackage{float}
\usepackage{graphicx}
\graphicspath{ {./} }

\begin{document}

\title[Trauma-Informed Social Media]{Trauma-Informed Social Media: Towards Solutions for Reducing and Healing Online Harm}


\author{Carol F. Scott}
\orcid{0000-0001-5811-0518}
\affiliation{%
  \institution{University of Michigan}
  \institution{School of Information}
  \city{Ann Arbor}
  \state{MI}
  \country{USA}
}
\email{cfscott@umich.edu}

\author{Gabriela Marcu}
\orcid{0000-0001-7150-5784}
\affiliation{%
  \institution{University of Michigan}
  \institution{School of Information}
  \city{Ann Arbor}
  \state{MI}
  \country{USA}
}
\email{gmarcu@umich.edu}

\author{Riana Elyse Anderson}
\orcid{0000-0002-5645-1738}
\affiliation{%
  \institution{University of Michigan}
  \institution{School of Public Health}
  \institution{Stanford University}
  \institution{Center for Advanced Study in the Behavioral Sciences}
  \city{Detroit}
  \state{MI}
  \country{USA}
}
\email{rianae@umich.edu}

\author{Mark W. Newman}
\orcid{0000-0001-7186-1383}
\affiliation{%
  \institution{University of Michigan}
  \institution{School of Information}
  \city{Ann Arbor}
  \state{MI}
  \country{USA}
}
\email{mwnewman@umich.edu}

\author{Sarita Schoenebeck}
\authornote{This project is based upon work supported by the National Science Foundation (NSF) under Grant 1552503.}
\orcid{0000-0002-8688-1595}
\affiliation{%
  \institution{University of Michigan}
  \institution{School of Information}
  \city{Ann Arbor}
  \state{MI}
  \country{USA}
}
\email{yardi@umich.edu}

\renewcommand{\shortauthors}{Scott, et al.}

\begin{abstract}
Social media platforms exacerbate trauma, and many users experience various forms of trauma unique to them (e.g., doxxing and swatting). Trauma is the psychological and physical response to experiencing a deeply disturbing event. Platforms' failures to address trauma threaten users' well-being globally, especially amongst minoritized groups. Platform policies also expose moderators and designers to trauma through content they must engage with as part of their jobs (e.g., child sexual abuse). We consider how a trauma-informed approach might help address or decrease the likelihood of (re)experiencing trauma online. A trauma-informed approach to social media recognizes that everyone likely has a trauma history and that trauma is experienced at the individual, secondary, collective, and cultural levels. This paper proceeds by detailing trauma and its impacts. We then describe how the six trauma-informed principles can be applied to social media design, content moderation, and companies. We conclude by offering recommendations that balance platform responsibility and accountability with well-being and healing for all.
\end{abstract}




\maketitle

\section{Introduction}
Trauma is a global health epidemic. According to the World Health Organization, 70\% of global people experience at least one traumatic event in their lifetime, and 30.5\% report four or more \cite{benjet2016epidemiology, kessler2017trauma}. In the United States (U.S.), estimates suggest that more than two-thirds of children report experiencing at least one traumatic event by age 16 \cite{copeland2007traumatic}. Approximately 90\% of adults experience some form of trauma at least once in their lifetime \cite{kilpatrick2013national,kessler2017trauma}, but experiencing multiple traumas is the norm, with three being the modal number of events per U.S. adult \cite{kilpatrick2013national, kessler2017trauma}. 

Trauma is much more than simply going through something that is ‘very stressful.’ It is the psychological and physical response to experiencing a deeply distressing or disturbing event, including all the ways that the trauma experience changes current and future behavior and well-being \cite{posner2008quality, dietkus2022call, itticmanual}. There can be long-lasting effects after experiencing even a single trauma (e.g., mugging), let alone multiple, repeated, or enduring events (e.g., racial trauma or the COVID-19 pandemic) \cite{abuse2014samhsa}. Re-traumatization is “a situation, attitude, interaction, or environment that replicates the events or dynamics of the original trauma and triggers the overwhelming feelings and reactions associated with them” \cite{AnnaIstitute2009}. It occurs when a person is reminded of the original traumatic event, causing them to experience the trauma all over again as if it was happening at that exact moment \cite{protocol2014trauma, AnnaIstitute2009, itticmanual}. Re-traumatization can be obvious (or not so obvious), is usually unintentional, but is always hurtful because it exacerbates the individual, group, or community's struggle and symptoms \cite{AnnaIstitute2009}.

At first, it might seem like trauma is a far-reaching topic for HCI. But, in 2022, Chen and colleagues published a CHI paper wherein they discussed how trauma-informed approaches could be applied to four broad areas of computing: user experience and research, design, security and privacy, and artificial intelligence and machine learning \cite{chen2022trauma}. Although they sparked and demonstrated collective interest in developing HCI perspectives on trauma-informed computing, many HCI researchers and practitioners have long been studying and working in trauma-related spaces. For example, a great deal of HCI research and practice focuses on understanding and mitigating various forms of abuse, such as intimate partner violence \cite{freed2017digital, leitao2018digital, tseng2021digital} and child sexual abuse \cite{bursztein2019rethinking, sultana_unmochon_2021, levine2020forensically}. Other HCI researchers study experiences that relate to trauma, including incarceration \cite{owens2021you, seo2021informal, ogbonnaya2018returning}, mental health concerns \cite{10.1145/3524017, nepal2021current, tsugawa2015recognizing}, postpartum loss and grief \cite{de2014characterizing, de2013predicting}, and work-related stress and burnout \cite{bell2003organizational, puspasari2019factors, armstrong2011challenges}. Similarly, many HCI researchers and practitioners are interested in mitigating harm by designing products and services that improve the lives of users, such as: understanding how trans people build community, find support, and express themselves online \cite{haimson2015disclosure, haimson2020trans}; understanding how sex workers use technology and how it could be an artifact for social justice \cite{sambasivan2011designing, strohmayer2019technologies}; and understanding how HCI researchers and designers can best support racially marginalized groups to make sense of and seek support on social media \cite{to2020they, burleson2020exploration}. 

Much less has been studied about how trauma manifests on social media, especially by the HCI community. Some social media-based traumas can be isolated and acute (e.g., sexual harassment via direct messaging between two people) while others can be large-scale, massive, and/or collective experiences (e.g., hateful subreddits or traumatic live and unfiltered images or videos shared globally) \cite{malby2015study, chandrasekharan_you_2017}. We conceptualize social media broadly---to include a wide range of platforms, products, and services, from the biggest and most popular ones that tend to come to mind first (e.g., YouTube, Facebook, Instagram, TikTok, Snapchat, and Twitter) to smaller ones and those to come \cite{Sinders2022}. We also see social media as an unsettled design space that ebbs and flows \cite{Sinders2022}---new platforms are developed almost every day, and popular ones become old or outdated (e.g., how teens think ‘Facebook is for [older] adults’ \cite{Sweney2018}).  

Historically, social media-based traumas were perceived as fringe experiences, lurking in the corners of a nascent Internet, but they are now mainstream \cite{vogels_state_2021,blackwell_harassment_2019}. Around the world, many people, groups, communities, and cultures can and do experience widespread harm on social media, which can be experienced as traumatic, and are increasingly documented \cite{goldberg_nobodys_2019,york2021silicon,amnesty_international_amnesty_2018,unesco_unesco_2021}. For example, recent research suggests that as many as 1 in 3 survey respondents across Australia, New Zealand, and the U.K. have experienced some form of image-based sexual abuse, which can be traumatic \cite{powell2022multi}. In the U.S., 41\% of adults say they have personally experienced some form of online harm, and 28\% report experiencing multiple and severe forms, including stalking, sexual harassment, and physical threats \cite{vogels_state_2021}. Many people also experience racist hate speech, unwanted misogynistic sexual content, ableism, stalking, doxxing, name-calling, misinformation, insults, impersonation, and public shaming \cite{vogels_state_2021,amnesty_international_amnesty_2018,gray_black_2020,goldberg_nobodys_2019,chandrasekharan_you_2017,bailey_misogynoir_2021,jackson_hashtagactivism_2020}. Of those who reported experiencing any harm online, most (75\%) say that their most recent experience occurred on social media \cite{vogels_state_2021, powell2022multi}. These abusive and traumatic experiences can have toxic effects, putting social media users at increased risk for adverse effects on their mental health (e.g., depression, anxiety, suicide, and eating disorders) \cite{vannucci2019social} or behavioral health (e.g., alcohol and substance misuse, self-harm, and chronic health conditions) \cite{tao2022exposure}. Traumatic experiences can also threaten communities’ ability to participate in equitable and democratic societies, both online and offline, \cite{york2021silicon} and lead to secondary or vicarious trauma for designers and content moderators \cite{hirsch2020practicing}. Re-traumatization on social media can be particularly concerning because posted content essentially lives forever, and algorithmic amplification can resurface traumatic content (e.g., a year-in-review reminder of a deceased child \cite{facebookapologies2014, Slate2014}).

One way to address social media-based trauma is by employing a trauma-informed approach. A trauma-informed approach is a relatively novel methodology in HCI. It involves applying trauma theory and principles (see Section 4.1) to the design and evaluation of interactive systems---the design and moderation of social media as well as to the social media companies themselves (i.e., their mission, policies, values, hiring practices, leadership styles, etc.). As we outline in Section \ref{sec:socialmedia}, a trauma-informed \textit{approach} stems from trauma-informed \textit{care}. Trauma-informed care has long been used in clinical practice by social workers, medical practitioners, and psychologists (e.g., trauma-informed cognitive behavioral therapy for depression) \cite{itticmanual, protocol2014trauma}. Trauma-informed design is applying a trauma-informed approach to the design of technology, spaces, and built environments \cite{jewkes2019designing, frey2020trauma, coloradotrust_using_2020, harte_rochetr, marcu2022toward}. Similarly, trauma-informed content moderation applies a trauma-informed approach to content moderation policies, procedures, and practices.

Applying a trauma-informed approach to social media---design, moderation, and companies---is timely and important, given the prevalence of online harm and that most of it occurs on social media \cite{vogels_state_2021, powell2022multi}. Trauma-informed social media design can complement other design approaches, such as human-centered design (HCD) or universal design. For example, while HCD centers people at all stages in the development process \cite{boy2009participatory}, not knowing enough (or anything) about trauma and, by extension, not being trauma-informed, can cause harm to anyone involved in the process, which not only includes users but also designers and moderators \cite{hirsch2020practicing}. Like universal design \cite{EUCouncil2007}, trauma-informed social media design attempts to create products, services, and systems that meet the needs of all users, not just the majority, as it assumes, more likely than not, all have a history of trauma. Even users who do not have trauma histories can benefit from this type of harm reduction or prevention \cite{dietkus2022call}. A trauma-informed approach can also guide standard social media design research methods, such as interviews. As Tad Hirsch (2020) explains, interviews can resemble therapy. When they are not trauma-informed, they can inadvertently cause harm not only to the participants but also to the design researchers \cite{hirsch2020practicing}. Further, asking people to share their experiences, thoughts, and feelings on sensitive topics as part of the design or interview process can be triggering and harmful to participants (who are asked to share) as well as the designers and researchers (who listen and receive the experiences shared) \cite{Wechsler2021, hirsch2020practicing}.

A trauma-informed approach to social media aligns with growing calls in HCI to acknowledge the lived experience and agency of users \cite{spiel2019agency} while also addressing systemic and structural problems rather than placing the burden on the individual \cite{ogbonnaya2020critical,bennett2019point,oliver2013social,walker2020more, harrington2020forgotten}. A trauma-informed approach to social media is also strengths-based \cite{rashid2009strength}, which means rather than focusing on deficits, it takes time to validate peoples' coping strategies, strengths, and capacities while recognizing that they are active participants in broader online systems \cite{rashid2009strength, harrington2020forgotten}. Strength-based approaches like this help to answer the ongoing HCI call for our community to co-design \textit{with} instead of \textit{for} users \cite{erete2018intersectional, dillahunt2017reflections}. Finally, a trauma-informed approach is harm-reductionist. In clinical addiction practice, harm reduction theory emphasizes minimizing the harmful consequences of behavior like alcohol abuse (e.g., reducing the number of hangovers or missed work) rather than eliminating the drinking. Harm reduction clinicians view any positive change---any reduction in related harm(s)---as having a meaningful impact on people's health and well-being \cite{leslie2008harm, single1996harm}. Individuals who seek this type of treatment are not interested in stopping their behavior but rather doing it in a safer or less harmful way \cite{leslie2008harm, single1996harm}. In applying this theory to social media use, we do not argue against its use altogether, we instead discuss how social media use can be less harmful. Any reduction in social media-based harm (i.e., less social media-based trauma experiences or re-traumatization) can have a meaningful impact on the well-being of users as well as designers and moderators who interact with traumatic content as part of their work \cite{saraiya2020perspectives}.

The overarching goals of this paper are to (1) synthesize the literature on trauma and trauma-informed approaches to provide a lens for the HCI community to evaluate aspects of social media design and moderation, and (2) generate new concepts and practices that meet business, user, and community needs without triggering people's existing wounds or creating new ones. By offering this unified lens, we provide a shared vocabulary, a collection of sensitizing concepts, and a mapping from social science theory and mental health practice to social media that can be a source for generating new design and moderation concepts and practices (e.g., social media company policies and procedures). In generating our analyses and recommendations, we draw upon the notion of “sensitizing concepts.” Sensitizing concepts originated in Grounded Theory and refer to how we can “draw attention to important features of social interaction and provide guidelines for research in specific settings” \cite{sas2014generating, bowen2006grounded}. Sas et al., argue that sensitizing concepts play an important role in HCI as a common form of “design implications” that “offer a preferred form of generalized knowledge for moving beyond the situatedness of requirements” \cite{sas2014generating}. Further, they “tend to state technology goals generated to meet specific social needs” \cite{sas2014generating} and therefore provide a mapping from social phenomena and their associated theories to technology design considerations. The contribution of this paper is a mapping between the experience of trauma and social media design approaches that might exacerbate or ameliorate such experiences.

To achieve these goals, we start by providing an overview of trauma (offline and online) and what it might mean if social media design, moderation, and relevant companies were trauma-informed. In applying a trauma-informed approach to the most prominent components of the social media ecosystem----social media design, moderation, and companies---we suggest what currently used concepts, methods, and design strategies to pay attention to, as they are or could be trauma-informed (e.g., just-in-time nudges), and which ones to avoid because they might not be trauma-informed and could therefore be more harmful than helpful (e.g., non-holistic content warnings). Each section of this paper builds upon the previous one. Sections 2 and 3 describe what trauma is, how trauma is experienced, and the impacts of trauma, both offline and on social media. In Section \ref{sec:socialmedia}, we introduce the concept of trauma-informed social media, providing a brief history and describing the six principles used to enact any trauma-informed approach. Drawing on and applying the large body of multidisciplinary trauma literature discussed in Sections 2-4, we propose trauma-informed social media design, moderation, and company strategies in Section 5. These recommendations are intended to help address complex platform ecosystems that enable trauma experiences, including design and policy decisions and the companies in which those decisions are made. As noted above, many HCI scholars and practitioners currently practice and design in ways that could be seen as trauma-informed. However, this approach is fragmented and, in some cases, incomplete. We aim to provide an organizing lens and discussion toward achieving trauma-informed social media. We also offer a few new recommendations that designers, moderators, and companies can add to their respective toolboxes and draw attention to the importance of engaging with trauma-informed approaches. Finally, in Section 6, we conclude with a reflection on the limitations of trauma-informed approaches and some potential hurdles to adopting its principles.

The contributions of this paper extend prior work in HCI, especially Chen et al. \cite{chen2022trauma}, in three overarching ways:
\begin{enumerate}
    \item Provide a richer overview of trauma and its impact, including an elaboration of the different types of trauma and how they can each can be experienced (Section 2).
    \item Extend and apply trauma theory and practice to social media, including how trauma manifests on social media and impacts users, moderators, and designers (Sections 3 \& 4).
    \item Offer various actionable sensitizing concepts, strategies, and recommendations to guide social media design, moderation, and companies (Section 5). While some of our recommendations are not completely novel, what is  novel is the trauma-informed lens we offer and apply in this context. 
\end{enumerate}

\subsection{Positionality Statement}
We draw on our expertise as social workers and psychologists with extensive trauma-informed clinical and research training, and as design and computing researchers with design and research experience. We have studied online harm in myriad contexts, which inspired our call for a trauma-informed approach to social media. While we spend considerable time thinking, reading, and talking about trauma, we acknowledge that this topic can be challenging for some and encourage readers to take care of themselves. 

\section{Background: Trauma \& Its Impact}
\subsection{What is Trauma?}
Trauma is much more than simply going through a ‘very stressful experience.’ It is about a life-changing event or experience that has a profound impact on the mind (emotionally and physiologically) and body (physically and physiologically) of those who experience it \cite{dietkus2022call, protocol2014trauma}. Rachael Diekus, a social work designer and trauma expert, offers a simplified and applicable definition of trauma: “Trauma is a response to anything overwhelming, and that happens too much, too fast, too soon, or for too long” \cite{dietkus2022call}. Understanding trauma importantly includes understanding how the individual, group, or community relates to it on a deeper level \cite{protocol2014trauma}, which we describe in the following subsection. 

While it can happen to anyone, trauma is more commonly experienced among individuals, groups, communities, and cultures who experience oppression and marginalization, including Black, Brown, and Indigenous people, People of Color; women and gender minorities; lesbian, gay, bisexual, queer people and sexual minorities (LGBTQ+ people); religious minorities; disabled people; and developmentally vulnerable populations (e.g., children and older adults) \cite{protocol2014trauma}. Compared to White Americans, Hispanic and Black Americans experience higher exposure to child maltreatment \cite{roberts2011race}. Black Americans also have significantly higher exposure to assaultive violence, such as kidnapping, mugging, and unwanted sex \cite{roberts2011race}. Globally, many refugees have survived traumatizing ordeals, including witnessing deaths by execution, starvation, beatings, and/or enduring violence and torture themselves \cite{steel2009association}. Trauma often occurs at an early age: In the U.S., 26\% of children witness at least one traumatic event before their fourth birthday; 1 in 5 children witness violence in their family or neighborhood; 60\% of youth have been exposed to crime, violence, and abuse, either directly or indirectly \cite{SAMHSA22, NorthDakota}; and Black youth report experiencing an average of five racial discrimination events daily, with most of them occurring on social media \cite{english2020daily}.

Broadly speaking, trauma can either be acute (a single event) or chronic (repeated and prolonged experiences) \cite{protocol2014trauma, itticmanual, bloom1999trauma}. But more precisely, a large body of multidisciplinary and global trauma research has identified 11 types of trauma. Although we review each of these specific types of traumas only once, they are not mutually exclusive, and the order of appearance does not denote a specific trauma’s importance or prevalence. We extend each of these 11 types to social media-based trauma in Section 2.

\begin{itemize}
    \item \textbf{Individual Trauma} mainly stems from a single event and typically occurs to one person (e.g., mugging or rape) \cite{amstadter2008emotional, barth2016military}.
    \item \textbf{Developmental Trauma} is an event that occurs during a specific period that influences later development or how one successfully ages (e.g., child sexual abuse, experiencing food or housing insecurity during childhood, or elder abuse) \cite{denton2017assessment, boullier2018adverse, protocol2014trauma, pereira2019elder}. 
    \item \textbf{Secondary \& Vicarious Trauma} are forms of trauma that people experience after witnessing trauma directly (e.g., seeing someone being assaulted) or hearing about it happening to another (indirectly) (e.g., hearing about someone being raped) \cite{dubberley2015making, guitar2017vicarious, protocol2014trauma}. Secondary trauma tends to occur suddenly, after directly or indirectly seeing or hearing about the trauma one time, whereas vicarious trauma develops over time through continual exposure to another person's, community's, group's, or culture's traumatic experience (e.g., when a clinician---therapist, doctor, or nurse---is exposed to prolonged patient suffering) \cite{guitar2017vicarious}.
    \item \textbf{Interpersonal Trauma} is often reoccurring (although it can be acute) and typically transpires between people who know each other, such as spouses, family, or friends (e.g., intimate partner violence [IPV] or elder abuse). \cite{protocol2014trauma,ernst2018trauma, pill2017trauma}.
    \item \textbf{Group, Community, or Collective Trauma} occurs when a particular group of people or community is traumatized (e.g., gangs who lose members from gun violence or firefighters who lose teammates in a fire) \cite{tosone2003shared, patton2017gang, protocol2014trauma}. These forms of trauma can affect a group's, community’s, or culture’s sense of safety (e.g., school shootings or disinvestment in specific neighborhoods) \cite{foster2001immigration, catani2008beyond, abuse2014samhsa, hirschberger2018collective}.
    \item \textbf{Racial Trauma} is caused by experiences of racial discrimination, threats of harm and injury, and humiliating and shaming events, which can begin at an early age. This can range from overt racism to microaggressions (e.g., police brutality against Black Americans or hate crimes committed against Asian Americans post-COVID-19 outbreak) \cite{comas2019racial, saleem2020addressing, protocol2014trauma, alvarez2016cost, anderson2019recasting, torino2018microaggression}.
    \item \textbf{Cultural Trauma} are events that erode the heritage of a culture with prejudice, disenfranchisement, and inequalities (e.g., unequal access to mental health service, education, and financial resources based on sociodemographic identity [e.g., income, race, age]) \cite{masiero2020individual, abuse2014samhsa, protocol2014trauma}.
    \item \textbf {Historical Trauma}, alternatively known as generational trauma, occurs when there is a historical event that is widespread and affects an entire culture (e.g., the Holocaust and slavery) \cite{kirmayer2014rethinking, brave2011historical, abuse2014samhsa, protocol2014trauma}. 
    \item \textbf{Natural Trauma} are sometimes colloquially thought of as ``acts of God''; they are typically unavoidable and are caused naturally (e.g., tsunami and avalanches) \cite{abuse2014samhsa, protocol2014trauma, Babbel2010}.
    \item \textbf{{Human-caused Trauma}} are those caused by human failures, including accidents (e.g., sports-related deaths), technological catastrophes (e.g., train derailment), and intentional acts (e.g., terrorism) \cite{abuse2014samhsa, protocol2014trauma}. 
    \item \textbf{\textbf{Mass Trauma}} can include disasters that impact a large number of people either directly or indirectly (e.g., the nuclear reactor meltdown in Ukraine or war) \cite{neria2011understanding, chrisman2014mass, protocol2014trauma}.
\end{itemize}

\subsection{The Adverse Effects on the Mind \& Body}
Many people, groups, and communities who experience trauma go on to live with little difficulty, and others experience lasting adverse effects \cite{protocol2014trauma, dietkus2022call, Chipman2022}. For the latter group, community, or individual, trauma can cause intense physical, emotional, psychological, and physiological harm, and be stored in the body and mind as pain, tension, or even numbness \cite{american2013diagnostic,protocol2014trauma, dietkus2022call, siegel2010mindsight}. David Trickey, UK psychologist and a representative of the UK Trauma Council, says that trauma is “a rupture in meaning-making....it forever changes the way individuals, communities, and groups see and interact with the world around them” \cite{Chipman2022}. Thus, for some, the impact of trauma can be profound and life-shaping. 

The adverse effects of trauma can be temporary or long-lasting; instant, prolonged, or delayed; subtle, gradual, or outright destructive; and emotional, physical, behavioral, and/or cognitive. Initial reactions typically include confusion, grief, sadness, anger, anxiety, nausea, elevated heart rate, restlessness, sleep and appetite disturbances, and difficulty concentrating \cite{abuse2014samhsa, dietkus2022call}. Delayed responses can surface days, weeks, or even months or years later, including persistent fatigue, sleep disorders, nightmares, depression, shame, and avoidance of activities associated with the trauma \cite{abuse2014samhsa}. More severe and persistent responses to trauma include continued distress and intrusive recollections of the event that disrupts the person's life, which are symptoms of the diagnostic criteria for post-traumatic stress disorder (PTSD) \cite{abuse2014samhsa, kessler2017trauma,breslau1999previous}. These changes can increase the risk of other diseases and chronic health conditions, including heart disease, cancer, and liver disease, to name a few \cite{bloom1999trauma}.

While trauma can impact those that experience it emotionally, physically, and cognitively, it is very contextual. How each individual, community, or culture responds will be different and based on various biological, psychological, social, and structural factors \cite{protocol2014trauma, dietkus2022call}. A person's sociodemographic identity (e.g., assigned sex, gender, age, race/ethnicity, and sexual orientation), history of trauma, capacity for coping, health, and mental health status all influence how they experience the trauma, the severity of the reaction that follows, and the long-term impact \cite{abuse2014samhsa}. For example, the Substance Abuse and Mental Health Services Administration (SAMHSA)---a leading public health agency within the U.S. Department of Health and Human Services---finds that women are twice as likely to experience post-traumatic stress disorder (PTSD) as a result of trauma compared to men; those who experience trauma during childhood and middle age are at greater risk for various mental and behavioral health adversities; Black Americans are exposed to trauma at higher rates compared to White Americans and are therefore more likely to experience PTSD; across all age groups, rates of trauma symptoms are higher among people who experience homelessness and are unsheltered; and LGBTQ+ people experience various forms of trauma (e.g., assault, bullying, and hate crimes) at disproportionately higher rates than their heterosexual peers do \cite{abuse2014samhsa}. Some regions and cultures are also more likely to experience a traumatic event or specific types of trauma, including those who experience military action and political violence, such as refugees and asylum seekers \cite{abuse2014samhsa}.

The experience of trauma tends to invoke two emotional reactions: either too much or too little emotion \cite{protocol2014trauma, dietkus2022call, siegel2010mindsight}. As such, many people, groups, or communities experience emotional dysfunction and can experience anger, anxiety, sadness, and shame \cite{protocol2014trauma}. Physically, those who experience trauma may report somatic complaints (e.g., pain, shortness of breath, fatigue), sleep disturbances, and gastrointestinal, cardiovascular, respiratory, and dermatological challenges, among others \cite{protocol2014trauma, dietkus2022call}. Without structural supports in place, those who experience trauma may engage in coping mechanisms to help manage the aftermath, including alcohol and substance use (self-medicating), self-harm, high-risk sexual behaviors, and over- or under-eating (compulsion) \cite{protocol2014trauma, howard2017childhood, madowitz2015relationship}. Cognitively, trauma can cause people to experience subsequent experiences---those similar to the trauma or even those that may seem far removed---as dangerous, experience excessive or inappropriate feelings of guilt, develop inaccurate rationalizations of the perpetrator's behavior, or develop intrusive thoughts or memories that come without warning \cite{protocol2014trauma, siegel2010mindsight}.

In the wake of trauma, the person, group, culture, or society must construct a new understanding of how the world works and how people exist in it \cite{harris2001using, dietkus2022call}. From the point of trauma onward, they construct a sense of self and others, as well as beliefs about the world, which may differ from those held prior \cite{harris2001using}. This sense-making then informs other life choices and guides the development of particular coping mechanisms \cite{harris2001using}.

\section{Social Media Trauma: Extending Trauma Theory \& Knowledge}
\subsection{What is Social Media-Based Trauma?}
We extend the 11 types of traumas and the impacts described above, providing some explanations and a few examples of how each can be experienced by social media users and vicariously by those who design and moderate social media. Again, these traumas are not mutually exclusive, nor do we provide an exhaustive list of all possible experiences. Nevertheless, what is described provides insight into some of the many ways trauma can be experienced on, or exacerbated by, social media and how a trauma-informed approach to design, moderation, and companies might help to decrease or eliminate these potentially life-altering events or experiences. 

\begin{itemize}
\item \textbf{Individual Trauma.} Online, one might experience individual trauma privately through harassment via direct message (e.g., on Instagram or Twitter) or publicly (e.g., posted on someone's timeline). These experiences can be traumatic, especially if the harassment is repeated or violent \cite{abuse2014samhsa}.
\item \textbf{Developmental Trauma.} Children can be exposed to age-inappropriate content online that may be traumatizing \cite{sun2021they,wisniewski2016dear}. Social media increases the opportunity or capacity for potential child sex offenders to gain access to children and child sexual abuse content \cite{malby2015study}. They also increase offenders’ pool of potential victims, provide them more opportunities to create false identities, and facilitate the sharing of harmful content with children \cite{malby2015study}. Through the use of social media, those who traffic humans can also recruit new victims, including children, and market child sex tourism \cite{malby2015study}. In the U.S., the Federal Bureau of Investigation (FBI) 2021 Elder Crime Report indicates that tech supported fraud is a growing problem that primarily targets people over 60, totally loses in the millions \cite{FBI2021}. Social media is a leading medium or tool facilitating these crimes \cite{FBI2021}.
\item \textbf{Secondary \& Vicarious Trauma.} If harassment is public, such as on Twitter or Reddit, it may cause secondary trauma to other social media users who witness it, especially among those who may have experienced similar traumas themselves. This can become vicarious trauma if users experience prolonged exposure. Social media designers and content moderators can also experience secondary or vicarious trauma. For example, moderators who repeatedly witness users' trauma through their shared content online (e.g., posts about self-harm, racial microaggressions, or child sexual abuse) may experience secondary or vicarious trauma. Similarly, design researchers may experience secondary or vicarious trauma from working with social media users---in reaction to the emotional demands of bearing witness to the lived experience of participants' trauma \cite{hirsch2020practicing}. Occupational stress and burnout are often the consequence of working in/for organizations that are not trauma-informed \cite{bell2003organizational}.  
\item \textbf{Interpersonal Trauma.} Similar to individual trauma, interpersonal trauma can be experienced through unwanted messaging on social media, which can be persistent, reoccurring, and hateful \cite{goldberg_nobodys_2019}. Social media also plays a powerful role in IPV, including providing new methods and opportunities for the abusive partner to have additional power and control, further isolate their target, and participate in sexual violence and stalking \cite{duerksen2019technological}. It is also a place where victim-blaming and humiliation can occur, which only add to the traumatic experience \cite{whiting2019online}.
\item \textbf{Group, Community, or Collective Trauma.} Group, community, or collective traumas can be experienced when entire groups are threatened online, such as misogynoir \cite{bailey_misogynoir_2021} or hateful subreddits like r/fatpeoplehate (which was eventually banned) \cite{chandrasekharan_you_2017}. Group trauma may also be experienced during networked harassment, when harassment campaigns are coordinated, long-term, and large-scale against groups of people \cite{marwick_morally_2021}. Further, with the proliferation of social media, people all over the world can now see vivid images of potentially traumatic events perpetually, sometimes live and unfiltered (e.g., the COVID-19 pandemic, U.S. 9/11, the U.S. murder of George Floyd, the 751 unmarked Indigenous graves found in Saskatchewan, Canada, ongoing honor killings in Pakistan, and the livestreaming of the terrorist attack in Christchurch, New Zealand). A growing body of literature suggests that this perpetual stream of content can have complex, unintended, and injurious impacts, especially for disenfranchised communities \cite{holman2020media,hopwood2017psychological,yeung2016roles}. In watching these events unfold and viewing these graphic images online, many people, communities, and collectives repeatedly endure trauma.
\item \textbf{Racial Trauma.} Racial trauma is experienced online via collective experiencing and sharing, and may often be re-traumatizing for months and years into the future, such as revisionist histories of the denials of racism in the murders of Breonna Taylor or George Floyd in the U.S. \cite{poynting2006tolerance,freelon2016beyond,jackson_hashtagactivism_2020, lamerichs2018elite, farkas2018cloaked}. The wide sharing of violence against Black people results in increased exposure to race-related traumatic events online \cite{maxie2022exposure}. Racist speech and behaviors, like misogynoir, which describes anti-Black and misogynistic behaviors, thrive online and are well-documented on a variety of online platforms \cite{bailey_misogynoir_2021,jackson_hashtagactivism_2020,chandrasekharan_you_2017,dosono2020decolonizing,patton2019sa}. 
\item \textbf{Cultural Trauma.} On social media, cultural trauma can occur when a culture (or members of a culture) experiences horrendous events, such as antisemitism and anti-Israel sentiments and posts (e.g., zionistagenda). Hashtags like zionistagenda regularly appear in conjunction with other harmful hashtags (e.g., devilworshipper and newworldorderagenda, israhell, and saturndeathcultkiller). These historic antisemitic tropes are reviving offline historical and mass traumas and retraumatizing these communities \cite{Antisemitism}. 
\item \textbf{Historical \& Broad-reaching Traumas.} We collapsed historical, natural, human-caused, and mass trauma types into an overarching category of historical and broad-reaching trauma since social media is too new to have known generational impacts. Nevertheless, we consider how these types of traumas might transpire on social media, especially in discussing or reliving offline traumas online. For example, most social media experiences of disasters have been discussions \emph{about} the disasters, which can include sense-making, mitigation, recovery, and misinformation that often occur in real-time and can induce heightened anxiety, sensitivity, and other reactions \cite{huang2015connected,starbird2011voluntweeters}. Also, the loss of access to social media itself---whether through government restrictions in some countries (e.g., \cite{york2021silicon}) or through infrastructure disasters due to extreme climate changes \cite{jyothi2021solar} or the digital divide \cite{Charkravorti2021}---could introduce newer kinds of trauma or exacerbate others. People globally rely on social media for connection and news, so instability could threaten supports. It can also be traumatic for users to see unfiltered and live streaming of natural, human-caused, or mass trauma as they unfold (e.g., tsunamis, terrorism, or war).
\end{itemize}

\subsection{Impact of Social Media-Based Trauma}
Social media platforms have distinct affordances related to trauma. For example, the persistence of certain content could re-traumatize viewers. Instead of an event happening once and then moving into memory, it can persist in perpetuity online. Similarly, algorithmic amplification means that potentially traumatizing content could be resurfaced repeatedly to users, further traumatizing or re-traumatizing them, such as the year-in-review reminder of a loved one who just passed \cite{morrison2020questions} or U.S. 9/11 terrorist attacks \cite{downs2016black}. 

Trauma literature also suggests that re-traumatization can ensue when a person, group, or community has to continually tell their story, is treated as a number, is seen as a label and not a person first (e.g., “addict”), have their trust violated, feels unseen or unheard, or has things done for them rather than with them \cite{harris2001using}. These characteristics are all built into social media platforms, which have relied on scale and automation rather than fostering healing and well-being. Asking people to continually tell their story can also be part of some HCI design and research methodologies. For example, interview studies that are interested in learning more about users' experiences with social media to inform design decisions \cite{hirsch2020practicing} (e.g., the recent well-being interventions on Instagram for people who experience depression, suicide, or eating disorders); more likely than not, they have a trauma history and interviewing them without employing a trauma-informed lens dramatically increases the risk of (re)traumatization, causing even more wounds and harm \cite{hirsch2020practicing}. 

Although harassment can be instantiated online, targets of online harassment frequently report disruptions to their offline lives, including emotional and physical distress, changes to technology use or privacy behaviors, and increased safety and privacy concerns \cite{lenhart_online_2016,goldberg_nobodys_2019,sambasivan_they_2019}. In networked contexts, such repeated exposure can inflict long-term damage to an individual's reputation, comfort, or safety. Notably, online harassment has a chilling effect on future disclosures, and social media users censor themselves for fear of being harassed for what they say \cite{lenhart_online_2016}.

A myriad of high-profile cases on social media has demonstrated links between online harm and trauma outcomes, such as anxiety, depression,  suicide, and job loss. Examples include the cases of Justine Sacco, who lost her job because of massive online outrage and harassment after an inappropriate tweet \cite{ronson2015one}; Adria Richards \cite{cutler2013dongle}, who also lost her job after an online post; and Matthew Hendrick \cite{goldberg_nobodys_2019}, who was threatened with repeated sexual assault in his home due to catfishing (i.e., the process of luring someone into a relationship using a fictional online persona) \cite{simmons2020catfishing, reichart2017follow}. In all cases, while the original tweets/posts were harmful and potentially traumatic, so was the retaliating harm---harm caused other damage and trauma by internet mobs. Gamergate is another example of people fearing their safety due to internet mob action \cite{Romano2021}. Recent research finds that Black youth who report experiencing more racial discrimination and traumatic events online also report higher trauma symptoms, including feeling worried about their safety and future, alienation, uncontrollable distress, and hyperarousal (i.e., when a person's body suddenly kicks into high alert) \cite{maxie2022exposure}. Some types of online harassment specifically aim to disrupt a person's offline life, such as doxxing (i.e., publicly sharing information about a person's home to encourage others to harass that person at their home) or swatting (i.e., falsely reporting a crime to encourage law enforcement to descend upon a person's home).

Online abuse can also result in fear for one's physical safety, regardless of whether or not threats of physical harm ever materialize. Revealing a person's home address, for example, results in a loss of perceived security that persists even after any online harassment has ceased. Further, social media-based trauma can result in physical reactions by exacerbating the bodily trauma of the original trauma (e.g., rape) with additional online trauma (e.g., sexual harassment) \cite{pater2016hunger}. People who experience trauma online may decide to leave social media to avoid re-traumatization, even though this means they may be isolated from the support networks on those sites \cite{ellison2011connection}. At an extreme, voices are silenced through threats of, or actual, violence and death. Qandeel Baloch, a social media icon in Pakistan, received abuse and harassment for her online persona that resulted in her brother murdering her as an “honor killing” \cite{reuters_brother_2019}. In South Korea, celebrities Hara Goo and Jin-ri Choi died by suicide, which many attributed to the networked harassment and abuse they experienced online \cite{sang-hun_south_2021}.

\section{From Trauma-Informed Care to Trauma-Informed Social Media}\label{sec:socialmedia}
A trauma-informed approach to social media stems from trauma-informed care. Trauma-informed care is an approach to patient care and clinical practice that “understands and considers the pervasive nature of all types of trauma and promotes environments of healing and recovery rather than practices and services that may inadvertently re-traumatize” \cite{itticmanual}. Simply put, trauma-informed care considers traumatic experiences when diagnosing, treating, and helping patients \cite{levenson_trauma-informed_2017, saraiya2020perspectives}. It has been at the core of social work's approaches to practice for over 50 years and highly influential in medicine and psychology \cite{levenson_trauma-informed_2017,raja_trauma_2015, Curi2018, saraiya2020perspectives}. Research consistently finds that trauma-informed care increases revenue (e.g., more service users/people come to treatment), cuts costs (e.g., decreases unnecessary treatment utilization), and improves patient outcomes (e.g., patient engagement, treatment adherence, health outcomes) \cite{menschner2016key, itticmanual}. Trauma-informed care also increases staff satisfaction, retention, and organizational commitment while decreasing staff burnout, stress, and fatigue \cite{itticmanual}.

Trauma-informed approaches have begun to gain traction in various contexts outside of patient and clinical care, including educational settings \cite{carello2015practicing}, penitentiaries \cite{adams2020trauma}, and website design \cite{Eggleston2017}. They have also been applied within interior design and architecture practices. For example, a Colorado supportive housing organization deliberately designed their facility to be trauma-informed by implementing specific design choices in their built space, including eliminating unlit stairwells, providing residents with beds that are made of natural materials (as opposed to metal frames), and using welcoming and calming colors and lighting. \cite{coloradotrust_using_2020}. They found that people felt “secure and relaxed,” like they “have a little bit more confidence and control,” and “feel like a person” \cite{coloradotrust_using_2020}. Trauma-informed approaches are beginning to be applied in HCI arenas, such as with Chen and colleagues’ 2022 paper and industry practitioners' work (e.g., see \cite{Eggleston2017, Wechsler2021, TiDSociety}). We extend these trajectories and work and contribute new insights to the HCI community by calling for trauma-informed social media and providing recommendations for social media design, moderation policies, and companies.

As noted in Section 1, the trauma-informed lens put forth in this paper offers the HCI community a shared vocabulary, a collection of sensitizing concepts, new ways of thinking, and design and moderation insights to consider. This paper also highlights what to pay attention to and why particular design and content moderation strategies matter, both new and existing. As with any sensitizing concepts and application of social science theory \cite{sas2014generating}, we stop short of providing specific concrete designs or guidelines, as the exact decisions will vary based on specific scopes of practice and business needs. Instead, we provide the HCI community with an important and helpful lens that has to been found to be very effective offline, in clinical settings \cite{menschner2016key, itticmanual}, and will likely be as effective online and in HCI practice.

\subsection{Toward Trauma-Informed Social Media: Getting Started by Applying the 6 Principles} 
There is no single path to becoming trauma-informed. To be trauma-informed exists on a spectrum, from being ‘trauma unaware’ to proactively and intentionally being ‘trauma-responsive’ and then ‘trauma-informed’ \cite{MissouriModel, Mass2020}. An important first step is to become ‘trauma-aware,’ wherein one understands what trauma is, the impacts trauma can have on those who experience it, and acknowledges that, more likely than not, we all have some lived experience of trauma. Even if one is not directly impacted by trauma, one undoubtedly knows someone who has been or is currently struggling with trauma. This is why trauma-informed approaches do \textit{not} require screening for or treating trauma, as working from the assumption that most people likely have a trauma history negates the need to screen. As a second step, because knowledge informs our insights and practices, it is imperative to take steps towards ‘trauma-sensitivity’ (e.g., have a working knowledge of trauma language and prevalence) and account for the potential presence of trauma in HCI work. In doing so, the HCI community can begin to examine how and why their social media design, moderation policies, and companies can be more trauma-informed. They can also become ‘trauma-responsive’ by operationalizing the tenets of a trauma-informed theory into concrete items, such as those we outline below. Designers, moderators, and companies can also begin to explore and understand the core principles and how they apply to their respective scopes of practice. Finally, being ‘trauma-informed’ happens when trauma theory and principles become ingrained and thoroughly embedded in one’s work.

\begin{figure*}
  \centering
  \includegraphics[width=\textwidth, height=8cm]{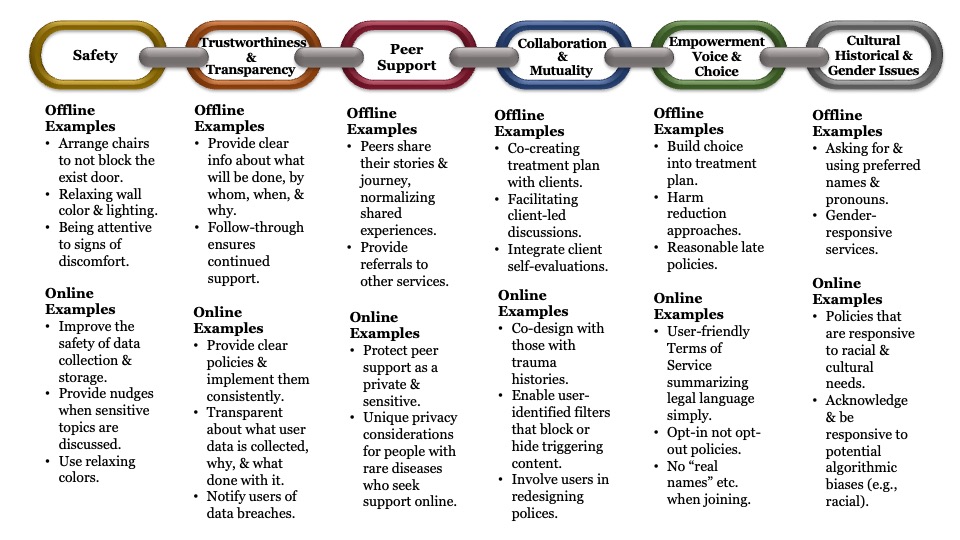}
  \caption{We apply the 6 guiding principles \cite{CDC2020, protocol2014trauma, abuse2014samhsa} to social media design, moderation, and companies and provide a few examples of how each principle can be enacted offline and online.}
  \Description{Depicts the 6 trauma-informed principles and applies them to social media. This figure also provides a few examples of how each principle can be enacted offline and online.}
    \label{fig:6principles}
\end{figure*}

All trauma-informed approaches are enacted through six guiding principles---safety; trustworthiness and transparency; peer support; collaboration and mutuality; empowerment, voice, and choice; and cultural, historical, and gender issues---which attempt to reduce the likelihood of re-traumatization \cite{coloradotrust_using_2020, itticmanual, protocol2014trauma, abuse2014samhsa, chen2022trauma, dietkus2022call, CDC2020} (see Figure \ref{fig:6principles}).\footnote{This figure is adapted from the SAMHSA and the CDC \cite{CDC2020}.} These principles are the same as those noted by Chen et al. \cite{chen2022trauma}, which SAMHSA established \cite{protocol2014trauma}, and a large body of multidisciplinary research has extensively examined across contexts (e.g., see \cite{decandia2014trauma, levenson_trauma-informed_2017, levenson2014incorporating}).\footnote{Unlike Chen and colleagues, we kept the original wording of all six principles used in clinical care ethics and standards (where the principles originated).} 

The six principles should \textit{not} be understood as a checklist but rather as sensitizing concepts (i.e., the background ideas or constructs that inform, highlight, and draw attention to important features and provide guidelines for design, research, and practice in specific settings \cite{bowen2006grounded, sas2014generating}). They are the ‘ingredients’ in a trauma-informed approach and are in place of a standardized or prescribed set of procedures because flexibility is key. Importantly, this flexibility ensures that they are generalizable across various settings and can be interpreted and applied in ways that are appropriate for those setting \cite{protocol2014trauma, itticmanual, coloradotrust_using_2020, abuse2014samhsa}. While we address each principle one by one, they are inherently interrelated, build on each other, and are not intended to be applied in a unidirectional way \cite{itticmanual, hales2017exploring} (refer to the interconnected chain in Figure \ref{fig:6principles}). Adopting a trauma-informed approach to social media is not accomplished through a single or particular technique. It requires constant attention, caring awareness, sensitivity, and, ideally, company-wide cultural change.

\begin{enumerate}
    \item \textbf{Safety: Physical \& Emotional}. \textbf{Physical Safety} involves the space in which interactions occur, including the consideration of security and aesthetics, such as appearance, wall colors (online or in-person), and accessibility \cite{harris2001using, protocol2014trauma}. \textbf{Emotional Safety} is being attentive to signs of discomfort, checking in, debriefing, providing support, and ensuring interactions are welcoming, respectful, and engaging \cite{harris2001using, protocol2014trauma}. Both must be considered when designing and moderating systems that are likely to serve people, especially those with trauma histories. 
    \item \textbf{Trustworthiness \& Transparency} involves providing clear information about what will be done, by whom, when, why, and under what circumstances (e.g., with collected user data). This principle also prioritizes privacy, confidentiality, respect, consistency, and continuity \cite{harris2001using, protocol2014trauma, dietkus2022call}. To be trustworthy, one must maintain respectful and appropriate boundaries and ensure interactions and rules are consistent and clear \cite{harris2001using, dietkus2022call}. 
    \item \textbf{Peer Support}. Peers are those with lived experience of trauma, or for children, a parent or guardian who cares for the child who experienced trauma and is subsequently vital to their healing process \cite{abuse2014samhsa, dietkus2022call, blanch2012engaging}. In online spaces, peers can offer support by sharing stories, disclosing recovery journeys, and linking and normalizing their shared experiences. 
    \item \textbf{Collaboration \& Mutuality} is about giving people options, working or co-designing with them, and providing them with clear and appropriate directives about their rights and responsibilities (e.g., including users in design and governance decisions, especially those with trauma histories or giving people an option about how their data are shared) \cite{harris2001using, itticmanual}.
    \item \textbf{Empowerment, Voice, \& Choice} similarly involves eliciting perspectives from the person who experiences trauma, asking them to determine their capacities and strengths, not simply giving the answers \cite{itticmanual}. Empowerment occurs when we focus is on peoples' strengths and help them find solutions \cite{harris2001using}. Collaboration refers to “the creation of an environment of doing with, rather than doing to or for someone” \cite{itticmanual}. In other words, people who experience trauma should be a part of shaping how they interact with platforms and online services. 
    \item \textbf{Cultural, Historical, \& Gender Issues} could be broadened to include any intersecting identities and experiences that shape how a person might experience trauma and what may be needed to support them. For example, for people who experience racism and ableism, providing trauma-informed design and governance would require recognizing how power differentials and structural inequalities reinforce their discrimination and harm experienced on- and offline \cite{harris2001using, dietkus2022call}. While this is a distinct and purposeful principle, diversity, equity, inclusion, and accessibility underscore the entire trauma-informed approach, including each of the preceding five principles; to be trauma-informed, one must be inclusive, equitable, just, and intentional \cite{itticmanual}.    
\end{enumerate}

In applying these six principles, trauma-informed social media designers, moderators, and companies prioritize the following four commitments \cite{protocol2014trauma}: \textit{Realizes} the impact that trauma has on users, employees, and those around them, \textit{Recognizes} the signs of trauma when they surface, \textit{Responds} to trauma with appropriate action, in a way that \textit{Resists Re-traumatization}. Next, we consider how the six principles of a trauma-informed approach can be incorporated into social media design, moderation policies, and companies. 

\section{Trauma-Informed Social Media Design, Moderation, \& Companies: Some Strategies \& Recommendations}\label{sec:strategies}
In this section, we apply the six principles (see Figure \ref{fig:6principles}) to address the various types of social media-based trauma described in Section 2. We focus on three of the most prominent components of the platform ecosystem: design, moderation, and companies. We offer the following recommendations as starting points because we know that sensitizing concepts can be notoriously difficult to translate into specific design implications \cite{sas2014generating}. We also offer some specific steps social media companies take toward becoming trauma-informed to decrease the ambiguity and potential difficulty of translating social science theory into action \cite{sas2014generating}. By invoking the notion of sensitizing concepts \cite{sas2014generating}, we also ask social media designers, moderators, and companies to reflect on how they might implement this unifying lens into their scopes of practice and support this advancement in design practice.

\subsection{Trauma-Informed Social Media Design}
Platforms continue to make design choices that push responsibility for harmful behavior onto its users, despite detailed documentation of this being problematic for over 30 years from early Internet scholars \cite{dibbell1994rape, daniels2009rethinking, daniels2015my, nakamura2014will, hughey2013racist}. For example, in 2022, Meta's VR “bubbles” create digital four-foot bubbles to prevent harassment, perpetuating a culture in which abusive behavior becomes the responsibility of the victim \cite{DailyBeast}. While the design examples and recommendations offered in this section are not an exhaustive list, they provide some insights into how designers can begin to apply trauma-informed principles in their work. Some of these strategies may not be new or novel---many are likely being used by designers across various HCI domain areas---but what is new or novel is the trauma-informed lens in this context. We aim to show the HCI community how what they already do may or may not be trauma-informed (e.g., how we can use nudges in a trauma-informed way or how content warnings must be part of a holistic, trauma-informed approach) and provide them with a few new ideas yet to be considered (e.g., use natural language processing to help users revise the wording in their posts and direct messages to be more trauma-informed).

\subsubsection{Content Warnings: Helpful or Harmful?}
‘Content warnings’ are one of the most widely used, yet controversial, aids to warn people about potentially upsetting, sensitive, or traumatic content---to prevent \textbf{\textit{secondary or vicarious trauma}}. They are used across multiple sectors, including in film, classroom material, books, museums, and social media \cite{charles2022typology, bridgland2022curiosity}. There is much disagreement about whether content warnings, trigger warnings, and content notes are the same, different, or parts of each other, but most use the terms interchangeably \cite{charles2022typology}. Further, some social media companies sometimes label it ‘sensitive content screens’ \cite{Twitter, bridgland2022curiosity} or trigger warning \cite{haimson2020trans}. For ease, we use the term content warning, but we intend for it to include all possible connotations.

Recent research, including systematic reviews, has explored the effectiveness of content warnings, and again, divergence emerges; some studies suggest they are appropriate and supportive \cite{byron2017infantilizing, lockhart2016trigger, haimson2020trans, stratta2020automated}, while more studies find that they have no effect or are even harmful, as they can be disempowering, victimizing, and marginalizing \cite{bryce2022pulling, jones2020helping, bridgland2022curiosity, ross2018fake, charles2022typology}. When they are “employed in isolation or in a tokenistic way, as a disingenuous gesture of trauma awareness,” Bryce et al. suggest that they can be particularly harmful because they “potentially inflame existing stressors and exacerbate maladaptive behaviors” \cite{bryce2022pulling}. Content warnings can also be counterproductive for those with severe trauma histories. Sometimes, as a protective mechanism, people who experience trauma deny the trauma (referred to as trauma denial), especially when they experience very traumatic events, such as rape (\textbf{\textit{individual trauma}}) or childhood sexual abuse (\textbf{\textit{developmental trauma}}) \cite{dietkus2022call}. Trauma denial is an innate process (i.e., the brain does this, and it is not a conscious decision) to protect the individual from emotional, mental, and physical pain \cite{church2017childhood}. Thus, while well-intentioned, content warnings can resurface a trauma the person has forgotten but now has no choice but to confront. Imagine, for example, this person is alone and scrolling through social media, and they encounter a content warning that triggers or invokes their denied trauma; they now must process this (re)traumatization on their own. Confronting the traumatic event and what it meant to them (alone) will likely bring up hurtful memories and sensations and could cause long-lasting harm \cite{church2017childhood}.

Although this area of research is far from settled, across the literature, an important take-home message is that for content warnings to be a valuable tool for assisting with the effective reduction of (re)traumatization (e.g., \textbf{\textit{secondary or vicarious trauma}}, \textbf{\textit{racial trauma}}, or \textbf{\textit{individual trauma}}), they must be embedded within a broader and holistic trauma-informed approach---like the one offered in this paper \cite{bryce2022pulling, charles2022typology, wyatt2016ethics, haimson2020trans}. Once part of a holistic, trauma-informed approach, they can improve \textbf{\textit{safety}}, \textbf{\textit{trustworthiness}}, and \textbf{\textit{choice}} while recognizing \textbf{\textit{cultural, historical, and gender issues}}.

\subsubsection{Tracking \& Responding to Harmful Experiences}
Social media platforms could detect and monitor users' trauma experiences; this may be especially important for those who experience chronic, networked harms, such as coordinated, long-term, and large-scale networked trauma (e.g., chronic \textit{\textbf{individual trauma}} as well \textit{\textbf{racial}}, \textit{\textbf{cultural}}, \textit{\textbf{collective traumas}}) \cite{marwick_morally_2021}. Tracking and detection tools could better support people who may be experiencing any form of trauma online by providing them with escalated protections to their accounts, such as user-initiated and controlled mass blocking features and reporting tools, as well as tools users manage to govern audiences better, including who can interact with those who have a trauma history. Partnering HCD with trauma-informed design, designers could enable users to determine what type of content is traumatic for them. For example, like Facebook or Google account privacy checkups, as users sign up for or continue to use a social media platform, they could answer some questions about potentially triggering, harmful, sensitive, or traumatic content (e.g., check boxes for all that apply), without having to explain why or provide any more detail. Users can also say how often, if at all, they would like to see content they self-identify as traumatic. Asking these types of well-being check-in questions to all users across all platforms is trauma-informed because it provides everyone, regardless of identity, location, type of platform used, etc., with \textbf{\textit{empowerment, voice, and choice}} and helps establish \textit{\textbf{physical and emotional safety}}. It is also equitable and inclusive, validates users' strength and resilience, and, importantly, prevents those who have a trauma history from (being the only ones) having to self-disclose, which, as discussed above, can be very harmful. To determine what content might be traumatic (when creating the well-being check-in questions), designers could refer to the types of trauma described above. They can also look to standardized trauma measures used by clinicians in clinical practice to screen for trauma (e.g., the Brief Trauma Questionnaire \cite{schnurr2002trauma}, the Life Events Checklist, DSM5 \cite{weathers2013life}, or the Adverse Childhood Experiences [ACE] \cite{felitti1998relationship}). They could also partner with trauma experts and co-design them.

One recent study suggests that the top 1\% of accounts on Facebook were responsible for 35\% of all observed interactions \cite{TheAtlantic2022}. To address this “Superuser-Supremacy Problem” \cite{TheAtlantic2022} and create a safer online environment for everyone, a trauma-informed approach focuses on helping those being harmed. Looking to tracking tools like Block Party App and their “Lockout Filters” \cite{BlockPartyApp} for direction and ideas can be helpful because they give users the power and agency to mass respond to harm (to reduce re-traumatization or trauma experiences like \textbf{\textit{individual}} or \textbf{\textit{racial trauma}}) and protect themselves from \textbf{\textit{generational}}, \textbf{\textit{historical}}, or \textbf{\textit{mass traumas}} that may be repeatedly shared online. By extension, they help achieve the principles of \textit{\textbf{safety}}; \textit{\textbf{choice, voice, and empowerment}}; and \textit{\textbf{cultural, historical, and gender issues}}.

\subsubsection{Providing Just-in-Time Nudges \& Text-Based Reinforcement Learning}
Applying and extending prior research \cite{wang2013privacy, sharma2020computational, sharma2021towards}, users could also be made aware or reminded of the impact they can have on (re)traumatizing others who may interact with their content (i.e., address the potential for \textit{\textbf{secondary or vicarious trauma}}). To do so, platforms could provide just-in-time (gentle) nudges that offer trauma-informed guidelines to users before they post or share content online, including pictures, time, and sentiment nudges \cite{wang2013privacy}. Many users do not think about nor remember who all their online friends and followers are \cite{wang2011regretted} and may not consider the potential impact of the content they share. Nudges could also help users consider how their post might be perceived (e.g., “this post might be perceived as traumatic, or this post may land racially traumatic for your Black friends”). Third-party software, such as browser plugins, could embed this trauma-informed design strategy. Just-in-nudges are an excellent example of \textit{\textbf{peer support}} as well as \textit{\textbf{collaboration and mutuality}}. 

Designers could also use natural language processing (NLP) and deep reinforcement learning (RL) models \cite{sharma2020computational, sharma2021towards} to transform text-based posts and messages that are not trauma-informed to ones that are. For example, as Sharma and colleagues have done \cite{sharma2020computational, sharma2021towards} in mental health support spaces with empathetic rewriting, trauma-informed designers could use NLP and RL to help users rewrite their posts or direct messages at the sentence level (e.g., to be more trauma-informed try changing “x” to “y” or “eliminating the specific detail X”) to reduce the likelihood of other users' experiencing \textit{\textbf{secondary or vicarious trauma}}. Designers could work with trauma experts to understand this multi-dimensional sensitizing concept, including using their trauma screening measures (some are mentioned in 5.1.1) to help identify all types of trauma. Models and processes like this are a form of \textbf{\textit{peer support}}. They can also foster \textbf{\textit{empowerment, voice, and choice}} and \textbf{\textit{collaboration and mutuality}}.  

\subsubsection{Trauma-Informed Filtering Tools}
Many people share potentially traumatic experiences on social media, including sharing one’s lived experience of being unsheltered \cite{woelfer2011improving,ledantec2008designs}; undergoing gender transitions \cite{haimson2015disclosure}; and a variety of public health and political crises, such as hurricanes \cite{metaxa2018social}, wildfires \cite{shklovski2008finding}, war \cite{hourcade2011hci}, conflict \cite{kwon2012audience}, terrorist attacks \cite{lindgaard2009mobile}, and displacement \cite{noyman2017finding,talhouk2016syrian,fisher2016future}. While sharing these experiences can be therapeutic for the person, they can also be harmful to users, moderators, and designers to witness, especially if the exposure happens repeatedly and if they share a similar lived experience \cite{morrison2020questions,downs2016black, schopke2022volunteer}. 

While users can mute key terms (e.g., war or rape) or specific users (e.g., those that perpetuate \textit{\textbf{individual}}, \textit{\textbf{racial}}, or \textit{\textbf{group trauma}}), trauma-informed platforms could provide users with additional tools that allow them to better filter and curate around imagery and video associated with trauma. For example, to prevent \textit{\textbf{secondary or vicarious trauma}}, content curation tools could ask users to rank pieces of content by the level of or potential to be traumatic. Also, the tools that search the best live websites for the latest content (news, blog posts, videos, and images) could filter out content that users suggest is traumatic before sharing it with users. Finally, as we noted in 5.1.1, interactive boards, like those used by Pinterest, could allow users to set filters that prevent them from seeing content they say is traumatic to them. Each of these examples enact all 6 principles---they increase \textit{\textbf{safety}}, \textit{\textbf{trustworthiness}}, and \textit{\textbf{transparency}}, are \textit{\textbf{collaborative}}, provide \textit{\textbf{empowerment, voice, and choice}}, and consider \textit{\textbf{cultural, historical and gender issues}}.   

\subsubsection{Support Accountability \& Repair}
To reduce \textit{\textbf{secondary or vicarious trauma}} and prevent most other types of trauma (e.g., \textit{\textbf{individual}}, \textit{\textbf{interpersonal}}, \textit{\textbf{racial}}, or \textit{\textbf{collective}}), trauma-informed social media platforms could also develop pathways for accountability that recognize and repair harm. Like trauma-informed design, centering accountability and repair requires a shift toward the needs of those traumatized. To accomplish this, designers can invoke various justice models, such as those that draw on restorative justice to recognize harm, establish accountability for that harm, and establish an obligation to repair harm \cite{schoenebeck_drawing_2020, schoenebeck_youth_2021}. Accountability could also include apologies, mediation, education, or other kinds of prompts and nudges. It can also allow people who violate platform guidelines to express their intent not to cause further harm. Together, these approaches help achieve \textit{\textbf{collaboration and mutuality}} and \textit{\textbf{empowerment, voice, and choice}}. Expressing intent not to commit further harm, with subsequent follow-through, could also be a form of restoration. However, for people who intentionally harm others but who refuse to accept accountability, punitive models, like banning, may be needed.

\subsubsection{Commit to Long-Term Support of Well-being}
In the field of social work, to be \textit{\textbf{trustworthy}} and \textit{\textbf{supportive}}, trauma-informed care requires follow-through that ensures an individual continues to receive the support they need, beyond the conclusion of a particular interaction or program. Trauma-informed social media platforms could similarly commit to supporting long-term well-being (e.g., recognizing that women journalists in many countries are subject to extreme, severe harassment and need enhanced protections and policies). However, designers must undertake these commitments with recognition of the sixth principle---\textit{\textbf{structural, gender, and cultural inequities}}---which requires critically reflecting on privilege, power, and status in identifying and recognizing trauma experiences. For example, Facebook pivoted from considering insults against men and women as equivalent to including historical inequities against women in deciding what kinds of content violate their guidelines \cite{nieman_facebook_2020}. In the same way that providers offline must be trained to support a wide range of identities and experiences, social media companies, designers, and moderators must recognize structural inequities in online experiences and seek to remedy them. In so doing, they can attempt to prevent or reduce the risk of experiencing \textit{\textbf{collective}}, \textit{\textbf{racial}}, and \textit{\textbf{broad-reaching traumas}}.

\subsection{Trauma-Informed Content Moderation}
In response to social media-based trauma, platforms have generally adopted governance approaches that focus on detecting individual pieces of content that violate community guidelines and then selecting a remedy---often deleting the content and sometimes banning the user in the case of repeated offenses \cite{schoenebeck_drawing_2020,goldman_content_2021}. However, these remedies overlook the harms and traumas associated with abusive behavior \cite{schoenebeck_drawing_2020,bailey_misogynoir_2021}. Platforms have also overlooked harm to people in their content moderation practices, which detect and remove harmful content, but ignore the needs of those impacted by the harm \cite{schoenebeck_drawing_2020,bailey_misogynoir_2021}. In this subsection, we offer some ways that social media moderation policies could be trauma-informed---as places to start and introductory insights.

\subsubsection{Increase Transparency of Content Moderation}
To date, platform governance and related policy decisions have relied on obfuscated processes of content moderation that have little transparency to all involved parties---content is deleted without leaving any visible trace of its removal \cite{schoenebeck_drawing_2020}. Instead of deleting and thus burying trauma experiences, moderation policies could be trauma-informed by developing principles of \textit{\textbf{transparency}} that disclose why the content was taken down or left up and how much content is removed or not \cite{bradford_report_2019, sunshine2003role, york2021silicon}. The framework of procedural justice proposes decision-making processes that are more \textbf{\textit{transparent}}, as they can increase perceptions of fairness \cite{sunshine2003role}. However, procedural justice must also be paired with the sixth principle, which recognizes how people's \textit{\textbf{cultural and gender identities and past histories}} will shape what decision-making should look like for them. For example, guidelines that result in racist content being allowed, but anti-racist counterspeech being removed, can cause and magnify \textbf{\textit{racial}} and \textbf{\textit{intergenerational trauma}} \cite{marshall2021algorithmic}. Just as any clinician who is not trained in trauma-informed care cannot identify or understand these kinds of traumas, social media companies that are not trauma-informed may be similarly failing to recognize traumas and design procedures that appropriately address them---a problem that should be remedied. Platforms could also increase (and achieve the trauma-informed design principle of) \textit{\textbf{trustworthiness}} by providing more transparent information about why they are making the decisions they make (e.g., why some content is removed while other content is not) and what data they collect and why.

\subsubsection{Provide Context Moderation \& Content Moderation}
Currently, social media companies primarily evaluate potentially traumatic behavior at the content level---that is, content moderators are asked to consider the specific words used in a given post or comment, divorced from the context of the post. That means they do not consider who the poster or target audience is or their relationship. However, complex behaviors like online trauma cannot be understood separate from the inherently social context in which they occur. For example, Desmond Patton and colleagues find that gang members process their shared \textbf{\textit{group trauma}} on Twitter in an attempt to obtain \textbf{\textit{(peer) support}} \cite{patton2018accommodating,patton2019sa}. Before making content moderation decisions (e.g., to delete this content), it would be essential to work with this community (i.e., the gang members) to understand their culture and grief and then make co-decisions based on the knowledge gained. Also, while the core experience of online trauma can be universal across regions and cultures (e.g., “dick pics” are universally problematic), according to trauma theory, how people experience harm may vary by individual, context, and culture. Non-consensual sharing of intimate images is an intense invasion of privacy regardless of the target's location. Still, for women in Pakistan or Saudi Arabia, for example, an intimate image could bring shame to an entire family. Even physical abuse, creating additional consequences and intensifying an already acute (\textbf{\textit{individual}} or \textbf{\textit{collective}}) trauma \cite{sambasivan_they_2019}. To evaluate context, platforms, designers, and moderators must also work in local communities and cultures and consider their languages and norms when creating platform design and moderation policies. In working with communities, groups, and cultures to develop trauma-informed moderation policies, social media companies, and employees can achieve the trauma-informed principles of \textit{\textbf{trustworthiness and transparency}}; \textit{\textbf{collaboration and mutuality}}; and \textit{\textbf{empowerment, voice, and choice}}.

Social media platforms that are trauma-informed could better foster online \textit{\textbf{safety}}---\textbf{\textit{emotionally and physically safe experiences}}---by focusing on user behavior rather than individual pieces of content. Social media users who are repeatedly traumatized online (which can be inferred from manual reports, network structures, and automated content analysis) could have better protection features available to them than a typical user, such as user-controlled mass blocking (noted above) and/or access to human customer support. These additional options support social media user \textit{\textbf{empowerment, voice, and choice}} they also increase \textbf{\textit{trustworthiness}}.

\subsubsection{Supporting \& Protecting Moderators}
As we note above, many people share potentially traumatic experiences on social media, which content moderators must engage with as part of their jobs. Witnessing these experiences can be harmful to moderators, especially if the exposure happens repeatedly (can lead to \textbf{\textit{secondary or vicarious trauma}}) and/or share a similar lived experience(s) (can lead to re-traumatization) \cite{morrison2020questions,downs2016black, schopke2022volunteer}.  

To help content moderators cope and decrease their experienced burnout, stress, and psychological distress \cite{schopke2022volunteer}, social media companies can continue to limit or prevent exposure and enact \textbf{\textit{safety}} by implementing supervised machine learning, natural language processing, and interface design decisions in trauma-informed ways (e.g., those mentioned above, in Section 5.1) \cite{steiger2021psychological}. They can also offer trauma training to moderators, have dedicated and trauma-trained team members and leaders that can recognize and respond to early warning signs of trauma and/or re-traumatization, and share resources related to receiving support and mental health counseling, particularly from \textbf{\textit{secondary and vicarious trauma}} experts \cite{itticmanual}. These trauma-informed strategies---individually and collectively---can increase \textbf{\textit{trustworthiness}}, \textbf{\textit{support}}, and \textbf{\textit{safety}}. Content moderators---paid and volunteers---should also set (and be supported to develop) healthy boundaries (e.g., exposure limits---in quantity [time each day] and duration [total time doing this job]), plan and schedule regular self-care and well-being breaks, and be a source of \textbf{\textit{peer support}} for each \cite{steiger2021psychological, itticmanual}. Company-wide trauma-informed changes, which we discuss next, are also necessary for supporting and protecting content moderators \cite{itticmanual, dscout}.

\subsection{Trauma-Informed Social Media Companies}
Because all trauma-informed approaches aim to help address complex ecosystems that enable or exacerbate trauma experiences, the companies in which those decisions are made must also be trauma-informed \cite{itticmanual, protocol2014trauma}; there cannot be pockets of designers, moderators, or employees who are trauma-informed within a company that does not support the trauma-informed mission and goals. For trauma-informed design and moderation to be successful, trauma-informed values and practices must be present and practiced throughout a company (e.g., culture, mission, and leadership) \cite{itticmanual, protocol2014trauma, dscout}. Recent research by dscout and HmntyCntrd supports this call to action: For tech companies to design and govern products that proactively prevent harm, they must first attempt to improve their employees' experience by investing in healing, seeing trauma as a systematic issue (not just an individual problem), and prioritizing the adoption of this paradigm and culture shift \cite{dscout}.

Becoming a trauma-informed organization is akin to and partners well with diversity, equity, inclusion, and accessibility (DEIA) organizational change. Like trauma-informed strategic change, prior work has found that for DEIA practices and approaches to be successful, the corresponding workplace and culture must be entrenched in these values \cite{creary2021improving, kulkarni2018promoting, sax2017diversifying, dubow2013diversity}. When companies fully embrace DEIA values, performance improves, and employees stay \cite{Catayst}. If we apply this DEIA research and the growing body of trauma-informed organization outcome research from clinical settings noted above (e.g., decreases cost and increases revenue), we can envision many similar benefits to having trauma-informed social media organizations---for users, employees, and society. Specifically, we anticipate future research finding that trauma-informed social media companies experience improved healthcare priorities for users and communities, maintaining peoples' active and constant use, and better content moderator and worker wellness.

\begin{figure*}
  \centering
  \includegraphics[width=\textwidth, height=8cm]{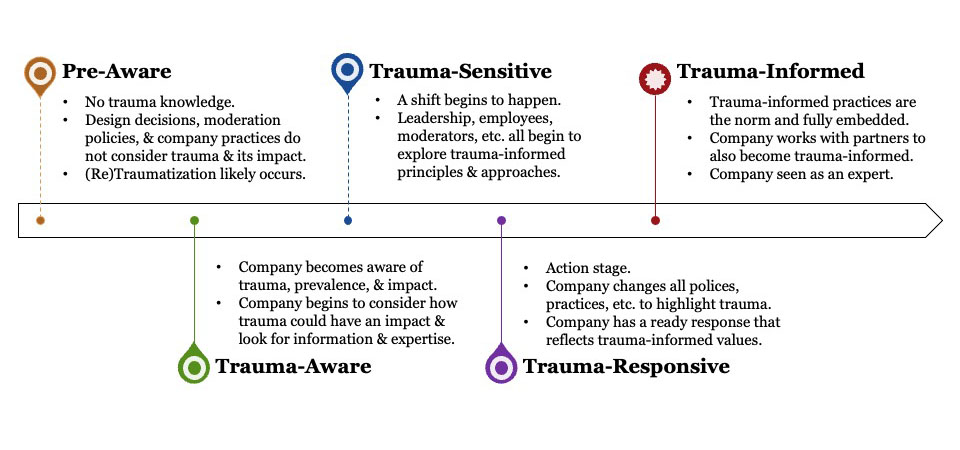} 
  \caption{For platforms to be trauma-informed, the larger companies must also be trauma-informed. To help social media companies think through and achieve this goal, we provide this adapted \textit{Missouri Model} \cite{MissouriModel} and describe it in more detail below. It is an iterative model that social media companies can apply over time.}
  \Description{Depicts the adapted Missouri Model for social media companies}
  \label{fig: fig2}
\end{figure*}

As a first step, to help social media companies think through and consider adopting a trauma-informed approach, we provide an adapted version of the “\textbf{\textit{Missouri Model}}” \cite{MissouriModel} (see Figure \ref{fig: fig2}). The state of Missouri Department of Mental Health and its partners developed an organizational change framework for implementing a trauma-informed approach within organizations. This model was designed to be applied to a wide range of settings \cite{MissouriModel}---from behavioral health settings to research departments, academic institutions, and social media companies. Notably, the Missouri Model acknowledges that change takes time; a company does not become trauma-informed overnight; it is a multi-year process \cite{MissouriModel}. Becoming a trauma-informed organization is iterative and on a continuum, wherein companies begin by becoming trauma-aware and conclude by maintaining trauma-informed.

By adapting, outlining, and applying this model, we aim to help social media companies understand, assess, and implement the basic principles of trauma-informed organizational change (across all the various settings within their respective organization) so that they can develop a framework that works best for them. Once conceptualized, because our adaptations are novel and have not yet been implemented nor tested within social media companies, future research can carry out our adapted Missouri model and provide some thick data on how an actual social media company grapples with the various stages of becoming trauma-informed. 

\subsubsection{Adapted Missouri Model: Outlining Some Steps \& Indicators of Success for Each Phase (refer to Figure \ref{fig: fig2})}
\paragraph{\textbf{Phase 1 \& 2: From Pre-Aware to Trauma-Aware}} Social media companies can begin the journey from being \textit{pre-aware} to \textit{trauma-aware} by learning what trauma is and understanding the value of being trauma-aware and informed can bring to their company. Some other steps a social media company can take to become trauma-aware include: (1) offer trauma awareness training to all employees; (2) help people within the company become aware of how and where to find additional trauma-related information and support employees in this learning; and (3) leadership can begin to explore what this new trauma-related information might mean for them and what steps they need to take next.  

\begin{itemize}
\item \textit{Indicators of being Trauma-Aware}: (1) most staff know what trauma is, are aware of the impact trauma can have, and they begin to change the way they see and interact with others; and (2) the impact of trauma is referenced in informal conversations among staff and leadership. 
\end{itemize}

\paragraph{\textbf{Phase 3: Trauma-Sensitive}} As social media companies move from \textit{trauma-aware} to \textit{trauma-sensitive}, they explore the principles and values of a trauma-informed approach, building consensus around how it applies to them and their work, and they prepare for change. Some specific steps can include: (1) conducting internal (company) self-assessments that identify existing strengths, resources, and barriers to change as well as current company practices that are both inconsistent and consistent with the values of trauma-informed design; (2) leadership can prepare the organization for change, including understanding staffs' readiness to change and identifying staff who can be change champions; and (3) start to hire people who reflect the values of trauma-informed design.

\begin{itemize}
\item \textit{Indicators of being Trauma-Sensitive}: (1) the company values and prioritizes the trauma lens, and a shift in perspective begin to happen; (2) trauma-informed values are identified in the mission statement and policy documents; (3) trauma training for all employees is institutionalized; (4) basic trauma information is available and visible to designer, moderators, and users via internal handouts, posts on the company's website and platform, part of job advertisements, etc.; (5) designers and moderators begin to seek out opportunities to learn new trauma-related skills; and (6) management recognizes and responds to trauma experienced by staff (e.g., secondary/vicarious). 
\end{itemize}

\paragraph{\textbf{Phase 4: Trauma-Responsive}} This is the action stage. Trauma-responsiveness happens when social media companies have begun to change their culture to highlight the role of trauma (at all organizational levels) and employees are rethinking their routines and infrastructure. To become trauma-responsive, social media companies can: (1) begin to integrate trauma principles into staff supports (e.g., addressing staff trauma, self-care routines, supervisions models, and staff development and performance evaluations); and (2) begin to integrate the trauma-informed principles into organizational structures, including record keeping, shared language across the company, and policies and procedures. 

\begin{itemize}
\item \textit{Indicators of being Trauma-Responsive}: (1) all staff apply trauma knowledge to their specific work; (2) language that supports the 6 principles is introduced throughout the company; (3) policies support addressing staff initial and secondary trauma; and (4) the company has a ready response for crisis management (e.g., platform and internal changes) that reflect trauma-informed values.  
\end{itemize}

\paragraph{\textbf{Phase 5: Trauma-Informed}} The final stage is achieved when the trauma-informed model has become so accepted and thoroughly embedded that it no longer depends on a few champions. Some specific steps to becoming and maintaining this stage include: (1) measuring the impact of this change on users, moderators, and designers; (2) company makes revisions to their trauma-informed policies and procedures, as needed; (3) company measures fidelity (adherence to this model), at least yearly; and (4) HR policies support hiring staff with knowledge and expertise (e.g., clinicians) in trauma then co-design with these individuals. 

\begin{itemize}
\item \textit{Indicators of being Trauma-Informed}: (1) leadership (existing and new hires) demonstrates a commitment to trauma-informed values and principles; (2) all staff is skilled in using trauma-informed design and governance; (3) all facets of the company have been reviewed and revised to reflect trauma-informed values; (4) people outside of the company (e.g., board of directors, users, and community members) understand the company’s mission to be trauma-informed; and (5) people turn to the social media company for trauma-informed expertise and leadership. 
\end{itemize}

\section{Trauma-Informed Approaches: Limitations, Critiques, \& Hurdles}
\subsection{Some Critiques \& Limitations}
Like any framework, a trauma-informed approach does not address all problems, nor is it free from potential problems of its own. For example, similar to most scholarly disciplines, the history and foundational theories of trauma studies “started and in some ways remain stuck in White, Euro-American frameworks” \cite{wright2021whitewashing}. However, scholars are working to uncover the voices of the many Black, Brown, and Indigenous people who played essential and contributing roles \cite{wright2021whitewashing}. Others suggest that the type of knowledge being shared in this paper---like informing others about collective forms of trauma (e.g., racial and historical)---is “a crucial first step in navigating the impact of historical oppression of Black, Brown, and Indigenous individuals and communities” \cite{GirlsInc}. There have also been revisions to trauma principles, including the addition of the sixth principles---cultural, historical, and gender---to bear witness to links between culture, identity, and trauma (see Figure \ref{fig:6principles}). This is also why we bring a broader lens of trauma theory to social media design and moderation, including that trauma experienced with and through social media can be collective and equally experienced by groups, communities, or cultures and not just individuals \cite{ginwright2018future, k2010shelter, abuse2014samhsa}. It is also why we suggest that social media companies must become trauma-informed. Finally, earlier framing of trauma theory and practices were critiqued for being too heavily rooted in neuroscience and medical models \cite{visser2015decolonizing, ginwright2018future}. Knowing this, we intentionally drew from a strength-based lens, not a deficit model. We also focused on improving, fostering, and strengthening user well-being, not reducing pathology. Finally, for social media, we believe a trauma-informed approach provides a new lens, highlights how some design strategies can be trauma-informed, and adds a new set of sensitizing concepts to design and moderation conversations (within and outside of the HCI community)---ones that may be incredibly generative for addressing large-scale social media-based trauma, can help meet the needs of users, and co-create safer environments where everyone is better heard.

\subsection{Potential Hurdles \& Solutions}
Facilitating complex change is never easy. This type of large-scale change can be logistically challenging and time-consuming, requiring resources (e.g., labor, funding, research). There might also be philosophical differences between sectors that need to work together, which can impede systemic change (e.g., different branches or sub-working groups within the larger company). One significant barrier could be the power imbalances and hierarchies within social media companies and between users and companies. Companies also have strong economic footholds in regulatory conversations, allowing them to control narratives about their policies and practices. Creating such change also requires a shift in perspective from reactive (worry about and solve problems as they come) to proactive (attempt to preclude harm and potential traumas from happening in the first place). Also, because trauma-informed social media is new, there is a lack of tools to measure organizational adherence or fidelity to the change model. Finally, we acknowledge that it can be difficult to measure organizational cultural change, sustainability of change, and whether any possible change (e.g., less harm to platform users and employees) is due to a trauma-informed company environment, change in platform design, or external pressures like regulation. Knowing that some of these hurdles exist, we suggest that future work could apply the adapted version of the Missouri Model \cite{MissouriModel} to social media companies (see Figure \ref{fig: fig2}). It is also why we outlined some specific steps social media companies can take and indicators of success at each stage. As an empirically-developed step-by-step guide for implementing and assessing organizational change, the Missouri Model acknowledges that this level of change happens over time and that missteps are likely to occur \cite{MissouriModel}. For example, similar to DEIA organizational change models, continuous efforts toward trauma-informed social media can bring new and fresh ideas into the workplace, improve organizational effectiveness, increase revenue, and strengthen user trust and loyalty.

For companies that decide to become trauma-informed, it is crucial that they look at the ‘big picture,’ noticing what they already have in place that could be or is trauma-informed (e.g., some of the current design strategies) and what they are missing or can do better, all while having the end goal in mind (i.e., achieving and sustaining stage 5 of the Missouri Model). To help assess progress along the way, companies can develop a checklist and corresponding Likert scale rating from 1=not yet started to 10=fully implemented and monitoring \cite{itticmanual}. To assess organizational culture, social media companies can use and adapt the Trauma-Informed Climate Scale \cite{hales2017exploring}, which measures employee perception of the company. The items are based on the six trauma-informed principles. To evaluate employee readiness to change, companies can use and apply the Transtheoretical Model, which posits that changes in behavior happen through six stages of change: pre-contemplation, contemplation, preparation, action, maintenance, and termination \cite{vickberg1999transtheoretical, grimley1994transtheoretical}. It also assumes that people do not change their behaviors quickly and decisively. Instead, change is ongoing and cyclical, especially when trying to change habitual behavior(s) \cite{grimley1994transtheoretical}. Specifically, companies can use the Transtheoretical model to measure success and determine the next steps, including assessing employees' current stage (i.e., which of the six stages they are currently at) and where they are throughout the change process. Finally, a large body of trauma-informed care literature has extensively examined its application and effectiveness across various social and human service agencies from which social media platforms can learn (e.g., see \cite{decandia2014trauma, domino2005service, k2010shelter}).

\section{Conclusion}
Around the world, social media users, especially dissidents, children, People of Color, refugees, and members of other non-dominant social groups disproportionately and frequently experience harm that can be traumatic (e.g., hate speech, non-consensual sharing of sexual photos, stalking, misinformation, impersonation, and public shaming). Those who experience these as traumatic must attempt to make sense of them and heal. Adding to their trauma experience, these types of harm are often overlooked, dismissed, or exacerbated by social media platforms that fail to account for or even acknowledge the systemic disparities that enable them. Also, moderators and designers must engage with traumatic content and lived experiences as part of their job, in addition to likely experiencing trauma themselves. But imagine a world where social media platforms, whether those currently in place or emerging models like public interest platforms, had a mandate of proactively promoting healing and well-being instead of reacting to harm as it happens. And imagine if social media design strategies and moderation policies more explicitly considered group, community, and societal inequities and trauma and, in doing, attempted to prevent harm or the likelihood of re-traumatization for all, including themselves.

Trauma-informed social media intends to not just react to the symptoms or issues related to trauma but to provide platforms that attempt to help users heal from these harms and prevent them from happening (again). A trauma-informed approach is an incredibly generative sensitizing concept for this; it highlights that offline and online experiences can span physical, psychological, historical, and collective trauma, leading to multiple and emerging kinds of harm in online environments. Expanding our scope of practice to include trauma-informed approaches can seem daunting and requires time, effort, support, and financial resources. However, as a source for generating design implications, trauma theory is a critical social science theory that the HCI community can and ought to apply \cite{sas2014generating}. We also believe it is work worth doing, as it can bring value to the social media platforms we design and moderate. Proactively engaging with the complexity of trauma is also worthwhile, as it better supports the well-being and healing of our users, participants, and employees and can improve the quality of our studies, products, and services. We also believe it could help social media companies answer the ongoing and growing call for social responsibility for the harm associated with using, designing, and moderating their products and services.

In applying the six trauma-informed principles to social media design, moderation, and company decisions, we contribute a new lens with new insights to the HCI community, especially since it can create more safe and more supportive social media experiences for all. This paper also supports and protects the well-being of content moderators, designers, and those around them. If we all become trauma-informed, we can move the field of social media toward a more mutually supportive, equitable, and healing environment.

\bibliographystyle{ACM-Reference-Format}
\bibliography{sample-base}

\end{document}